\documentclass[prl,aps,twocolumn,superscriptaddress,showpacs]{revtex4}
\usepackage{amssymb}
\usepackage{amsfonts}

\usepackage{graphicx,graphics,epsfig}
\usepackage{dcolumn}
\usepackage{bm}
\usepackage{amsmath}
\usepackage{verbatim}
\usepackage{color}
\usepackage{subfigure}
\usepackage{times,natbib}
\usepackage{amsmath,amsfonts,amssymb,graphics,graphicx,epsfig,color,times,natbib}

\begin{document}
\title{Quantumness of Product States}

\author{Jing-Ling Chen}
 \email{chenjl@nankai.edu.cn}
 \affiliation{Theoretical Physics Division, Chern Institute of Mathematics, Nankai University,
 Tianjin 300071, People's Republic of China}
 \affiliation{Centre for Quantum Technologies, National University of Singapore,
 3 Science Drive 2, Singapore 117543}

\author{Hong-Yi Su}
 \affiliation{Theoretical Physics Division, Chern Institute of Mathematics, Nankai University,
 Tianjin 300071, People's Republic of China}
 \affiliation{Centre for Quantum Technologies, National University of Singapore,
 3 Science Drive 2, Singapore 117543}
\author{Chunfeng Wu}
 \affiliation{Centre for Quantum Technologies, National University of Singapore,
 3 Science Drive 2, Singapore 117543}

\author{C. H. Oh}
\email{phyohch@nus.edu.sg}
\affiliation{Centre for Quantum
Technologies, National University of Singapore, 3 Science Drive 2,
Singapore 117543} \affiliation{Department of Physics, National
University of Singapore, 2 Science Drive 3, Singapore 117542}

\date{\today}

\begin{abstract}
Product states do not violate Bell inequalities. In this work, we
investigate the quantumness of product states by violating a certain
classical algebraic models.  Thus even for product states,
statistical predictions of quantum mechanics and classical theories
do not agree. An experiment protocol is proposed to reveal the
quantumness.
\end{abstract}


\maketitle

The physical picture in quantum mechanics dramatically differs from
that in classical local realistic theories. For instance, the
intriguing nonlocal correlation called entanglement predicted by
quantum mechanics has drawn many researchers' attention, and
fruitful applications have been achieved in quantum information
science~\cite{Nielsen}. Another significant property is nonlocality,
detected by the violation of Bell inequalities. Gisin
theorem~\cite{Gisin} states that any two-qubit entangled pure states
violates Bell inequality.
But in general these two concepts are different from one another,
and so far there has been no clear bound that separate the degree of
entanglement and nonlocality~\cite{Grothendieck}. Moreover, The
experiment tests for excluding local realistic theories have long
been suffering from simultaneously closing the loopholes of locality
and
detection~\cite{loophole1,loophole2,loophole3,loophole4,loophole5,loophole6}.

When the system $\rho$ is in the separable state, i.e., $\rho=\sum
p_i\rho^i_A\otimes\rho^i_B$, Bell inequalities will then not be
violated, which implies the possibility of simulating the system by
a certain hidden-variable local realistic theory. In 2002, Ollivier
and Zurek proposed quantum discord~\cite{discord} as a novel
measurement for multi-particle correlations. In their definition,
separable states have non-vanishing quantum discord, unless only one
component $\rho^i_A\otimes\rho^i_B$ remains. Throughout the paper,
this kind of state is referred to as a product state. Consequently,
a product state is usually considered as the classical state for
quite a long period.

In this work, surprisingly, we show in product states of
multi-particle systems ,there exists quantumness, defined by the
violation of Alicki-Van Ryn (AR) inequality~\cite{AR-I}. It seems
that quantumness is quite common in physical systems.

In algebraic model~\cite{AM}, observables are elements of a certain
$C^*$-algebra, and states are positive normalized functionals
$A\mapsto\langle A\rangle_{\rho}$ with $\langle A\rangle_{\rho}$
denoting the mean value of the observable $A$ in the state $\rho$.
In a classical algebraic model, any two elements $A$ and $B$ are
commutative, then one gets the following implication
\begin{eqnarray}
0\leq A\leq B\Rightarrow A^2\leq B^2.\label{classical}
\end{eqnarray}
The mean value of the observable $\langle A\rangle_{\rho}$ has two
definitions: (i)classically, $\langle A\rangle_{\rho}$ is defined by
$\int A(x)\rho(x)dx$ where $\rho(x)$ is some probability
distribution, and (ii)in quantum mechanics $\langle A\rangle_{\rho}$
is defined by $Tr(A\rho)$ where $\rho$ is the system state. In other
words, a classical model must satisfy the AR inequality
\begin{eqnarray}
\langle A\rangle\geq 0,\\
\langle B\rangle\geq 0,\\
\langle B-A\rangle\geq 0,\\
\langle B^2-A^2\rangle\geq 0.
\end{eqnarray}
However, in quantum mechanics there exist noncommutative observables
that violate the fourth constraint, namely, one can find
positive-definite observables $A$ and $B$ satisfying $\langle
A\rangle\geq 0,\langle B\rangle\geq 0,\langle B-A\rangle\geq 0$,
while $\langle B^2-A^2\rangle< 0$. This violation is called
quantumness. Experimental tests have been performed for the case of
one qubit.

For two-qubit product state
$|\phi\rangle=|0\rangle\otimes|0\rangle$, one can write down the
following positive-definite matrices $A$ and $B$:
\begin{eqnarray}
A&=&\left(\begin{matrix}1+\frac{\sqrt{295}}{40}&0&0&-\frac{27}{40}\\
0&1+\frac{\sqrt{759}}{160}&-\frac{25}{32}&0\\
0&-\frac{25}{32}&1-\frac{\sqrt{759}}{160}&0\\
-\frac{27}{40}&0&0&1-\frac{\sqrt{295}}{40}\end{matrix} \right),\\
B&=&\left(\begin{matrix}\frac{3}{2}&0&0&-\frac{9}{20}\\
0&\frac{5}{4}&-\frac{5}{8}&0\\
0&-\frac{5}{8}&\frac{5}{4}&0\\
-\frac{9}{20}&0&0&\frac{3}{2}\end{matrix} \right),\label{2-qubit}
\end{eqnarray}
so that
\begin{eqnarray}
\langle\phi|A|\phi\rangle&=&1+\frac{\sqrt{295}}{40}\geq 0,\\
\langle\phi|B|\phi\rangle&=&\frac{3}{2}\geq 0,\\
\langle\phi|(B-A)|\phi\rangle&=&\frac{20-\sqrt{295}}{40}\geq 0
\end{eqnarray}
hold, while
\begin{eqnarray}
\langle\phi|(B^2-A^2)|\phi\rangle=\frac{65-4\sqrt{295}}{80} <0,
\end{eqnarray}
indicating the quantumness of this two-qubit product state.


A proposed experiment protocol is to find a system whose Hamiltonian
is in the form of $B^2-A^2$, then measuring the ground-state energy
would reveal the quantumness. Any $4\times 4$ Hermitian operator can
be in the following form
\begin{eqnarray}
\mathcal {H}=\sum_{i,j=0}^3\beta_{ij}\sigma_i\otimes\sigma_j,
\end{eqnarray}
where $\sigma_0$ is unity identity, $\sigma_{1,2,3}$ is Pauli
operators, and coefficient $\beta_{ij}$ is defined by $Tr(\mathcal
{H}\sigma_i\otimes\sigma_j)$. If Hamiltonian $H=B^2-A^2$, for
instance, then the system state is
\begin{eqnarray}
\rho=\frac{e^{-H/kT}}{Tr(e^{-H/kT})}=\frac{e^{-E_n/kT}|n\rangle\langle
n|}{\sum_{n=1}^4(e^{-E_n/kT})},
\end{eqnarray}
with temperature $T$. The last step is justified by the fact that
$H$ is diagonal in our two-qubit case. When $T\rightarrow 0$, all
positive-energy weights $e^{-E_n/kT}$ vanish; and given the ratio of
energy of ground and excited states, only the weight for ground
state energy remains, hence $\rho\rightarrow |00\rangle\langle 00|$,
a nondegenerate pure state.

The generalization to arbitrary-qubit systems can be realized by
enlarging matrices $A$ and $B$. Firstly, one has to find several
groups of positive-definite $2\times 2$ matrices $A$ and $B$ that
violate the inequality (\ref{classical}). Secondly, insert the
matrix elements of $A_2$ inside $A_1$, that is,
\begin{eqnarray}
\left(
  \begin{array}{cccccccc}
       (A_1)_{11} &  &  &(A_1)_{12}  \\
       & (A_2)_{11} & (A_2)_{12} &    \\
       & (A_2)_{21} & (A_2)_{22} &  \\
      (A_1)_{21} &  &  & (A_1)_{22}  \\
     \end{array}
\right).
\end{eqnarray}
The rest entries are set to zero. Thus $A_1$ has been enlarged into
a $4\times 4$ matrix $A$; similarly, one can obtain the enlarged
$4\times 4$ matrix $B$, as shown in (\ref{2-qubit}). Here we suppose
the minimal eigenvalue of $B_1^2-A_1^2$ is less than that of
$B_2^2-A_2^2$ without generality. Thirdly, by repeating the second
step, one finally obtains the desired matrices that violate
(\ref{classical}).
\begin{eqnarray}
\left(
  \begin{array}{cccccccc}
   (A_1)_{11} &  &  &  &  &  &  & (A_1)_{12} \\
     & \ddots&  &  &  & &  & \\
     &  & (A_2)_{11} &  &  &(A_2)_{12} &  &  \\
     &  &  & (A_2)_{11} & (A_2)_{12} &  &  &  \\
     &  &  & (A_2)_{21} & (A_2)_{22} &  &  &  \\
     &  & (A_2)_{21} &  &  & (A_2)_{22} &  &  \\
     & \swarrow &  &  &  &  & \ddots &  \\
    (A_1)_{21} &  &  &  &  &  &  & (A_1)_{22} \\
  \end{array}
\right)
\end{eqnarray}
Note that the enlarging process has been performed in such a way
that the ground state is always $(1,0,0,...,0)^T$. Generally
speaking, $A$ and $B$ should be $2^n\times 2^n$ square matrices, and
at least two groups of $A$ and $B$ have to be found so as to acquire
the enlarged positive-definite matrices, since degeneracy is
permitted for excited states. Additionally, one may find other group
of positive-definite matrices whose minimal eigenvalue of $B^2-A^2$
is even less than $\frac{65-4\sqrt{295}}{80}$. Numerical results
show that the minimal eigenvalue could be $-0.059$. In our case, we
use two groups of $A$ and $B$, namely,
\begin{eqnarray}
A_1=\left(\begin{matrix}1+\frac{\sqrt{759}}{160}&-\frac{25}{32}\\
-\frac{25}{32}&1-\frac{\sqrt{759}}{160}\end{matrix} \right),\;\;
B_1=\left(\begin{matrix}\frac{5}{4}&-\frac{5}{8}\\
-\frac{5}{8}&\frac{5}{4}\end{matrix} \right),
\end{eqnarray}
and
\begin{eqnarray}
A_2=\left(\begin{matrix}1+\frac{\sqrt{295}}{40}&-\frac{27}{40}\\
-\frac{27}{40}&1-\frac{\sqrt{295}}{40}\\
\end{matrix} \right),\;\;
B_2=\left(\begin{matrix}
\frac{3}{2}&-\frac{9}{20}\\
-\frac{9}{20}&\frac{3}{2}\\
\end{matrix} \right),
\end{eqnarray}
with respective minimal eigenvalues $\frac{65-4\sqrt{295}}{80}$ and
$\frac{501-20\sqrt{759}}{1600}$ of $B^2-A^2$.

In summary, we have shown two-qubits product state exhibits
quantumness indicating the obvious deviation from classical models.
Then an experiment protocal has been proposed so as to reveal the
quantumness. The generalization to arbitrary-qubit cases has also
been investigated with the conclusion that quantumness is a common
property existing in physical systems in product states, contrary to
the preceding understanding of what classical states are, which
needs further investigation. The quantumness is based on the
violation of classical algebraic models, which is intrinsically
distinct from local-hidden-variable  models.

\begin{acknowledgments}
J.L.C. is supported by National Basic Research Program  (973
Program) of China under Grant No.\ 2012CB921900 and NSF of China
(Grant Nos.\ 10975075 and 11175089). This work is also partly
supported by National Research Foundation and Ministry of Education,
Singapore (Grant No.\ WBS: R-710-000-008-271).
\end{acknowledgments}

\end{document}